\documentclass[aps,prl,showkeys,showpacs,amssymb,cite,floatfix,
amsfonts,epsf,preprintnumbers,nofootinbib,superscriptaddress]{revtex4-1}

\usepackage[dvips]{graphicx}
\usepackage{bm,latexsym,amsmath,amssymb,amsfonts}
\usepackage[usenames,dvipsnames]{color}
\usepackage[colorlinks=true,linkcolor=blue]{hyperref}
\usepackage{color}
\usepackage{soul}
\usepackage{epsfig}
\usepackage{filecontents}
\newcommand{\newc}{\newcommand}
\newc{\beq}    {\begin{equation}}
\newc{\eeq}    {\end{equation}}
\newc{\beqa}    {\begin{eqnarray}}
\newc{\eeqa}    {\end{eqnarray}}
\newc{\bs}    {\section}
\newc{\no}    {\\ \nonumber}

\def\apj{{\em Astrophys. J.  }}
\def\apjl{{\em Astrophys. J. Lett. }}

\def\mnras{{ Mon. Not. Roy. Astron. Soc.  }}

\begin{document}
\title{Radial Acceleration Relation from Ultra-light Scalar Dark matter}

\author{Jae-Weon Lee}
\affiliation{Department of electrical and electronic engineering, Jungwon university,
            85 Munmu-ro, Goesan-eup, Goesan-gun, Chungcheongbuk-do,
              367-805, Korea}

\author{Hyeong-Chan Kim}
\affiliation{School of Liberal Arts and Sciences, Korea National University of Transportation, Chungju 27469, Korea}

\author{Jungjai Lee}
\affiliation{Division of Mathematics and Physics, Daejin University, Pocheon, Gyeonggi 487-711, Korea}
\affiliation{Korea Institute for Advanced Study 85 Hoegiro, Dongdaemun-Gu, Seoul 02455, Korea}

\begin{abstract}
We show that ultra-light scalar dark matter (fuzzy dark matter) in galaxies has a quantum mechanical
 typical acceleration scale about $10^{-10}\,\mbox{ms}^{-2}$,
which  leads to the baryonic Tully-Fisher relation. Baryonic matter at central parts of galaxies acts as a boundary condition for dark matter wave equation
and influences stellar rotation velocities in halos.  Without any modification of gravity or mechanics this model also explains
the radial acceleration relation and MOND-like behavior of gravitational acceleration
found in galaxies having flat rotation curves.
This analysis can be extended to the Faber-Jackson relation.
 \end{abstract}

\maketitle


The baryonic Tully-Fisher relation (BTFR)~\cite{McGaugh:2000sr}
is a tight  empirical correlation between the total baryonic mass ($M_b$) of a disk galaxy and its asymptotic rotation velocity $v_f$; $M_b\sim v_f^4$.
Semi-analytic models for BTFR  based on baryonic processes in a cold dark matter (CDM) cosmology predict significant scatter  from individual galaxy formation history, but observed BTFR  is largely independent of baryonic processes and has small scatter~\cite{2041-8205-816-1-L14}.
There is another strong relation  called radial acceleration relation (RAR)
 between the radial gravitational acceleration
traced by rotation curves (RCs) of galaxies and
predicted acceleration by the observed baryon distributions~\cite{PhysRevLett.117.201101}.
There are models~\cite{Ludlow:2016qzh} based on CDM paradigm explaining RAR, but it is unclear whether this tight relation can survive chaotic processes of galaxy formation and mergering.
These relations are puzzling, because galactic halos seem to be dark matter (DM) dominated objects and
RCs  at outer parts of galaxies are believed
to be governed mostly by DM not by baryons.
There are other  relations challenging conventional DM models such as
 Faber-Jackson relation or  baryon-halo conspiracy~\cite{Trippe:2014hja}.

On the other hand
BTFR and RAR are  consistent with Modified Newtonian dynamics (MOND) which was proposed
to explain the flat RCs without introducing dark matter~\cite{1983ApJ...270..365M}.
According to MOND Newtonian gravitational acceleration of baryonic matter $g_b$ should be replaced by
\beq
g_{obs}=\sqrt{g_b g^\dagger},
\label{MOND}
\eeq
when $g_b<g^\dagger\simeq 1.2\times 10^{-10}\mbox{ms}^{-2}$. The value of  $g^\dagger$ can be determined from RCs
of galaxies~\cite{1983ApJ...270..365M,Milgrom:1992hr}.
However, MOND also has its own difficulties in explaining the properties of galaxy clusters and cosmic background radiation~\cite{Dodelson:2011qv}.

 In this letter, we show that
 ultra-light scalar dark matter (fuzzy dark matter) has a quantum mechanical  typical acceleration scale $g^\dagger$,
which naturally leads to dynamically established BTFR.
Without any modification of gravity or mechanics this model also explains
the RAR and MOND-like behavior of gravitational acceleration.

Although the CDM model  well explains
observed large scale structures of the universe, it encounters
many difficulties in explaining galactic structures.
For example, numerical studies with CDM predict  cuspy DM halos
 and many satellite galaxies, which are in tension with  observational data~\cite{Salucci:2002nc,navarro-1996-462,deblok-2002,crisis}.
 Recently, there have been  renewed interests in
 scalar field dark matter ~\cite{1983PhLB..122..221B,1989PhRvA..39.4207M,sin1,myhalo,0264-9381-17-1-102}  (SFDM, often also called fuzzy DM ~\cite{Fuzzy}, ultra-light axion, BEC DM or wave DM)
 as a solution of these problems.
 In this model
  DM is  a ultra-light scalar with mass $m \simeq 10^{-22} e{\rm V}$ in Bose-Einstein condensation (BEC).
  Its long  Compton wavelength $\lambda_c=2\pi \hbar/mc\simeq 0.04 {\rm pc}$
   suppresses the  formation of  structures smaller than a galaxy, while
 it plays the role of  CDM at  super-galactic scales.
(See Refs. \citealp{Lee:2017qve,2011PhRvD..84d3531C,Hui:2016ltb,2014ASSP...38..107S,2014MPLA...2930002R, 2014PhRvD..89h4040H,2011PhRvD..84d3531C,2014IJMPA..2950074H,Marsh:2015xka}
 for a  review and references.)
 Since galaxies are non-relativistic objects, the typical length scale $\xi$ of a galaxy
  is about the de Broglie length $\xi_{dB}$ rather than $\lambda_c$, which helps in solving
  the problems of CDM.

  In this model, galactic halos are self-gravitating
 giant boson stars where gravitational force of matter balances
with quantum pressure from the uncertainty principle with spatial uncertainty  $\xi$ about $\xi_{dB}$.
 From the uncertainty principle $\xi m v\simeq \xi m \sqrt{GM_{c}/\xi}\ge \hbar$
  one can estimate $\xi\simeq \hbar^2/GM_{c}m^2$, where
 $M_{c}$ is the halo mass scale and $v$ is a typical rotation velocity of a galaxy. If we identify $M_{c} \sim 10^8 M_\odot$ and
  $\xi\sim 300 {\rm pc}$  to be the typical mass and the size of the core of a dwarf galaxy, then
   $m\simeq \hbar/\sqrt{\xi GM_{c}} \simeq O(10^{-22})e{\rm V}$. Note that $\xi$ is not a constant but
    almost independent
   of other properties of the galaxy except for $M_c$.
 We suggest that $\xi$ and the uncertainty principle lead to a natural acceleration scale $g^\dagger=GM_{c}/\xi^2\simeq \hbar^2/m^2\xi^3\simeq O(10^{-10}) \mbox{ms}^{-2}$
  of SFDM.
We will show that this acceleration scale $g^\dagger$ from the uncertainty principle  gives a hint
to the aforementioned relations of galaxies.

In SFDM  model, DM scalar field $\phi$
is described by the action
\beq
\label{action}
 S=\int \sqrt{-g} d^4x[\frac{-R}{16\pi G}
-\frac{g^{\mu\nu}} {2} \phi^*_{;\mu}\phi_{;\nu}
 -U(\phi)],
\eeq
where the typical potential is
$U(\phi)=\frac{m^2}{2}|\phi|^2+\frac{\lambda}{4}|\phi|^4$.
For fuzzy DM $\lambda=0$.
  In the Newtonian limit
  the Einstein equation and the Klein-Gordon equation from the action
  can be reduced to
the Schr\"{o}dinger equation  ~\cite{PhysRevD.35.3640}
\beq
\label{GPE}
i\hbar\partial_t \psi (\bold{r},t)=-\frac{\hbar^2}{2m}\nabla^2\psi(\bold{r},t)+ m \Phi\psi(\bold{r},t)
\eeq
and the Poisson equation
\beq
\Delta \Phi(\bold{r})=  4\pi G (\rho_d(\bold{r})+\rho_b(\bold{r}))
\eeq
with a self-gravitation potential $\Phi$ and wavefunction $\psi\equiv\sqrt{m}\phi$.
Here, $\rho_d$ is a DM density and $\rho_b$ is a baryonic matter density, both of which
 contribute to  $\Phi$.
 Since galaxies are non-relativistic, in this model a galactic DM halo is well
 described by the macroscopic wavefunction $\psi$
 which is a solution of the Schr\"{o}dinger equation.

For simplicity we consider a spherical
 fuzzy DM halos.
Integrating the
above equation gives magnitude of total gravitational acceleration
\beq
g_{obs}(r)\equiv |\nabla \Phi|
=\frac{4\pi G }{r^2} \int_0^r (\rho_d(r')+\rho_b(r')) r'^2 dr'\equiv  g_d(r) + g_b(r),
\label{phieq}
\eeq
where $g_d(r)$ is the acceleration from dark matter and $g_b(r)$ from baryonic matter
at galactocentric radius $r$.
The Madelung representation ~\cite{2011PhRvD..84d3531C,2014ASSP...38..107S}
\beq
\psi(r,t)=\sqrt{\rho_d(r,t)}e^{iS(r,t)/\hbar}
\label{madelung}
\eeq
is useful to calculate $g_{obs}$ in a fluid approach.
Substituting Eq. (\ref{madelung}) in to the Schr\"{o}dinger equation, one can obtain
a modified Euler equation
\beq
\frac{\partial \textbf{v}}{\partial t} + (\textbf{v}\cdot \nabla)\textbf{v} +\nabla \Phi
+\frac{\nabla p}{\rho_d} -\frac{\nabla Q}{m} =0,
\label{euler}
\eeq
where
  $\textbf{v}\equiv \nabla S/2m$, $p$, and $Q\equiv\frac{\hbar^2}{2m}\frac{\Delta \sqrt{\rho_d}}{\sqrt{\rho_d}}$ are a fluid velocity, the pressure from a self-interaction (if $\lambda \neq 0$), and a quantum potential, respectively.
The quantum pressure ${\nabla Q}/{m}$  helps fuzzy dark matter to overcome the small scale problems  of CDM
and plays an important role in this paper.

 By taking $\textbf{v}=0$ and $\partial_t \textbf{v}=0$, we find a stationary equilibrium condition
\beq
g_{obs}(r)=g_d(r)+g_b(r)
=\frac{\hbar^2}{2m^2} \left|\nabla\left(\frac{\Delta \sqrt{\rho_d}}{\sqrt{\rho_d}}\right)\right|,
\label{gobs}
\eeq
 which describes the dynamical balance between the gravitational attraction
and the quantum pressure.
This is the key equation to understand the origin of RAR in our model.
It is interesting that the fuzzy DM density profile $\rho_d$ and hence the wavefunction
$\psi$ traces
the total gravitational acceleration not just $g_d$.
Using an approximation $\partial_r\sim 1/\xi$ in $\nabla Q/m$
 one can define the
  characteristic acceleration for fuzzy DM halos more precisely
\beq
g^\dagger  \equiv \frac{\hbar^2}{2m^2 \xi^3}
=2.2\times 10^{-10} \left(\frac{10^{-22}e{\rm V}}{m}\right)^{2}
\left(\frac{300{\rm pc}}{\xi}\right)^3  \mbox{m/s}^2 .
\label{gdagger}
\eeq
  Note that this scale has a quantum mechanical origin
 which is a unique feature of fuzzy DM. $g^\dagger$ defined in this way is almost
  independent of $\rho_d$. This fact might explain the universality of $g^\dagger$.
  However, in realistic situations galaxies with different masses can have somewhat different $\xi$,
  and hence, $g^\dagger$ can have some ranges in our model.
  \begin{figure}[htbp]
\includegraphics[width=0.4\textwidth]{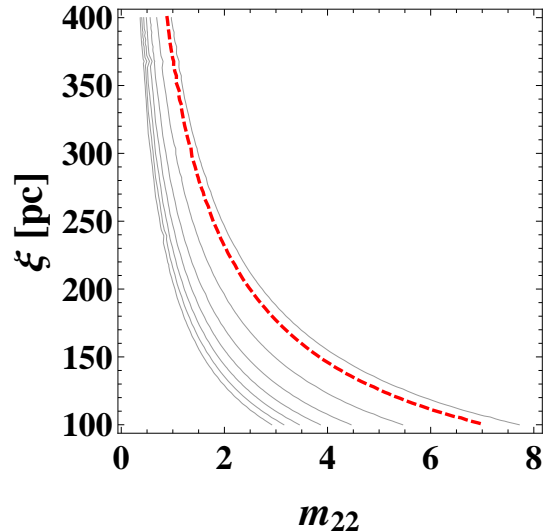}
\caption{(Color online) $g^\dagger$ as a function of $\xi$ and $m_{22}\equiv m/10^{-22}e\mbox{V}$.
The thin lines correspond to  $g^\dagger=(7,6,5,4,3,2,1)\times 10^{-10} \mbox{m}/\mbox{s}^2$
from the left, respectively.
The dashed red line represents the observed value $g^\dagger=1.2\times 10^{-10} \mbox{m/}\mbox{s}^2$.}
\label{rarfig1}
\end{figure}
 Quite interestingly, if we use the typical core size of the dwarf galaxies ($\sim  300\,\mbox{pc}$ ~\cite{Strigari:2008ib})
 as  $\xi$, one can reproduce the observed
  value $g^\dagger=1.2 \times 10^{-10} \mbox{m/}\mbox{s}^2$ for a favorable mass $m=1.35\times 10^{-22}e\mbox{V}$.
  Fig. \ref{rarfig1} shows an effect of the parameter $\xi$ on $g^\dagger$ for a given $m$.

\begin{figure}[htbp]
\includegraphics[width=0.4\textwidth]{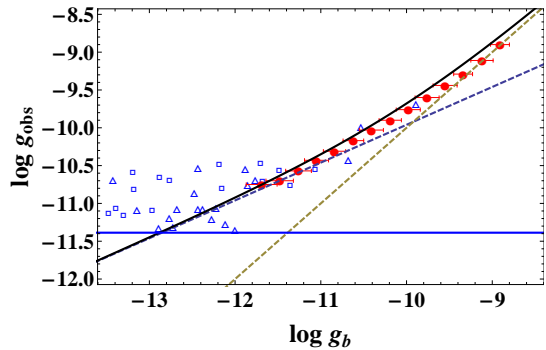}
\caption{ (Color online)
RAR between $g_b$ and the observed acceleration $g_{obs}$.
 The red dots represent the binned  values of 2693 points from
SPARC database~\cite{Lelli:2016zqa}. Triangles and boxes represent data
for dwarf spheroidal galaxies  extracted from Ref.
\citealp{2017ApJ...836..152L}. Two dashed lines are
1:1 line and $\sqrt{g^\dagger g_b}$ line with $m=1.35\times 10^{-22}e\mbox{V}$ and $\xi=300\, \mbox{pc}$, respectively.
The black solid curve represents our  theoretical approximation $g_{obs}\simeq g_b+\sqrt{g^\dagger g_b}$
with the same  $m$ and $\xi$.
The blue horizontal line represents the typical acceleration $g_0$ for dwarf galaxies with $M_{tot}=10^{8} M_\odot$.}
\label{rarfig}
\end{figure}

Let us see how $g^\dagger$ affects galaxies.
 According to precise numerical studies  with fuzzy DM~\cite{2014NatPh..10..496S}
 a massive galaxy has a soliton-like core with size $\xi=O(10^2)$ pc
 surrounded by a virialized halo of granules (also with size $\sim \xi$)
  having a Navarro-Frenk-White (NFW) density profile.
 In the regions where $g_{obs} \gg g^\dagger$ (as in a center of a galaxy)
 baryonic matter is usually more concentrated than fuzzy DM
 and
 the gravitational acceleration mainly comes from baryon mass.
 On the other hand, a DM dominated region at a large $r$ beyond the core
   usually has $g_{obs} \le g^\dagger$, because $g^\dagger$ represents the typical
 acceleration of DM cores if they were made of only fuzzy DM.
   Therefore, for massive galaxies, $g^\dagger$
   acts as a  parameter discriminating   baryonic matter dominated regions ($r \ll \xi$)
 from  DM dominated  regions ($r \gg \xi$).
For baryonic matter dominated regions such as central parts of massive galaxies
 $g_b  \gg g_d$, and obviously $g_{obs} \simeq g_b \gg g^\dagger$,
 which explains the 1:1 linear part of RAR graph in Fig.~\ref{rarfig}.

On the other hand, there are three regions where  $g_{obs}$ can be much smaller than $g^\dagger$;
I) Outermost edge of galaxies ($r> O(10^2) \mbox{kpc}$).
II) Outer parts of massive galaxies with almost flat RCs ($\mbox{kpc}<r<O(10) \mbox{kpc}$).
III) Small dwarf galaxies ($r<\mbox{kpc}$).

Unlike MOND, in our model if a galaxy is well isolated from others,
the rotation velocity in the region I
is expected to drop off because of lack of matter. For example, the Milky way and
 earlier galaxies seem to have
falling RCs~\cite{Bhattacharjee:2013exa,2017ApJ...840...92L}  in the outermost edge.
However, observational data in this region  is still rare and uncertain,
so we ignore this region in this letter to understand the observed RAR.

Since the observational data points satisfying Eq. (\ref{MOND}) mainly come from the region II,
and BTFR also relies on the flat rotation velocity data in this region, we will first focus on
the flat RCs for which $g_{obs}\sim r^{-1}$.
There are many attempts to obtain the flat RCs with SFDM using
excited states \cite{sin1,myhalo,Bar:2018acw} or specific potentials ~\cite{Schunck:1998nq,Guzman:1999ft}.
To find the RAR in the region II in fuzzy DM models we need to know $\rho_d$.
Numerical studies with only fuzzy DM
 indicate that DM halos
have a solitonic core with size about $\xi$ surrounded by an NFW-like profile
from virialized granules ~\cite{2014NatPh..10..496S}. Thus, an average DM density over the granules
 for this quasi-stationary system can be roughly given by using a step function $ \Theta $ ~\cite{Marsh:2015wka};
\beq
\rho_d(r)\simeq \Theta(r_e-r)\rho_{sol} + \Theta(r-r_e) \rho_{NFW}.
\label{rhomarsh}
\eeq
Here $\rho_{sol}\propto 1/(1+ (r/r_c)^2)^8$ is a soliton density, and the NFW profile is
$\rho_{NFW}= {\rho_{d0}}r_0/{r(1+r^2/r_0^2)}$
with  constants $\rho_{d0}$, $r_0$, $r_e$ and $r_c$.
Quite interestingly, however,
a recent numerical work~\cite{2018MNRAS.478.2686C} found that if we include  baryon (stars) in the inner halo,
the total matter density $\rho_{tot}\equiv \rho_b+\rho_d$ follows an almost
 isothermal profile ($\rho_{tot}\sim r^{-2}$ and $g_{obs}\sim r^{-1}$) near the half-light radius $r_h$ of the baryon matter
 rather than Eq. (\ref{rhomarsh}).
 The only cases exhibit this features are when $\rho_b$ is comparable to $\rho_d$
 at the half mass radius, which is consistent with the arguments about
 $g^\dagger$ below Eq. (\ref{gdagger}). That is, $r_h$ is the position where $g_{obs}\sim g^\dagger$ and
 the DM dominance and the flat RCs start. The physical origin of this numerical behavior is unclear, but it seems to be a kind of averaging effect
 of log-slope of DM density and baryon matter density ~\cite{2018MNRAS.478.2686C}.
 If we accept the numerical result, in fuzzy DM model,
 the region where $g_{obs}\ll g^\dagger$ in massive galaxies usually corresponds to the region with almost flat RCs
and $r<O(10) \mbox{kpc}$ as observed.

In this region, we can find a relation between $g_{obs}$ and $g_b$ by a simple reasoning.
 As $r$ increases beyond baryon dominated regions, $M_b(r)$ slowly approaches a total baryon mass $M_b={\rm const.}$,
  and
    $g_b(r)$  decreases faster than $g_d(r)$ does.
 At a point $r^\dagger$ the acceleration $g_b$ becomes comparable to $g_d$, and
$g_{obs}$  approaches the  typical value $g^\dagger$, which means  $g_b(r^\dagger)\simeq g^\dagger/2$
and RCs become flat. The above numerical work ~\cite{2018MNRAS.478.2686C} indicates that $r^\dagger$ is about
the half-light radius, i.e., $M_b(r^\dagger)\simeq M_b(r_h)=M_b/2$.
 Therefore, $2 g_b(r^\dagger)\simeq GM_b/r^{\dagger 2}\simeq g^\dagger$.
 From Eq. (\ref{gdagger})  it implies
 \beq
 r^\dagger\simeq\sqrt{GM_b/g^\dagger}=\frac{\sqrt{2 Gm^2\xi^3M_b}}{\hbar}.
 \eeq
  Thereby, a bigger $M_b$ means a larger $r^\dagger$.
 Around this point $g_{obs}=|d\Phi/dr|$ starts to be small and
 the rotation velocity graph $v(r)\simeq \sqrt{|\Phi(r)|}$ has a gentle slope,
 which means almost flat RCs, i.e., $v(r)\simeq v_f$ ~\cite{2018MNRAS.478.2686C}.
 Using $r^\dagger$ above one can estimate the constant rotation velocity
 \beq
  v_f\equiv\sqrt{r^\dagger g^\dagger}=\sqrt{\frac{GM_b}{r^\dagger}}\simeq (GM_b g^\dagger)^{1/4},
 \eeq
 which is just BTFR,
 $M_b=A v_f^4$
 with
 \begin{eqnarray}
 A&=&(G g^\dagger)^{-1}=\frac{2 m^2 \xi^3}{G \hbar^2} \nonumber \\
&=&34.16  \left(\frac{m}{10^{-22}e\mbox{V} }\right)^{2}
\left(\frac{\xi}{300\mbox{pc} }\right)^3  M_\odot /(\mbox{km/s})^4.
\label{BTF}
 \end{eqnarray}
 Remarkably, with Eq. (\ref{gdagger}) it reproduces the observed value $A=47\pm 6 ~M_\odot \mbox{km}^{-4}\mbox{s}^4$ ~\cite{1538-3881-143-2-40},
 if  $m=1.173\pm 0.07\times 10^{-22}e\mbox{V}$ for $\xi=300 \, \mbox{pc}$.
 Note that $r^\dagger\sim O($kpc$)$ is somewhat larger than $\xi$ for a typical galaxy.
 One of the advantages of our approach is that approximate values of $g^\dagger$ and $A$ can be derived
 from the model.
 In our model, BTFR has a quantum mechanical origin, although it is a relation
among macroscopic quantities of baryonic matter.
 (A  Tully-Fisher-like relation between the total DM and the circular velocity
was suggested for fuzzy DM in Ref. \citealp{Bray:2014dca}.)
 Following  Ref. \citealp{2018arXiv180301849W}
  we can derive the asymptotic form of RAR
  from the BTFR ($M_b=v^4_f/Gg^\dagger$),
 \beq
 g_b(r\gg r^\dagger)\simeq \frac{G M_b}{r^2}=\frac{1}{g^\dagger}\left(\frac{v_f^2}{r}\right)^2
   =\frac{g_{obs}^2}{g^\dagger},
   \label{gb}
   \eeq
   i.e., $g_{obs}=\sqrt{g_b g^\dagger}$.
 This is  the MOND-like behavior of $g_{obs}$ in the RAR graph
 at large radii
  where $g_b \ll g^\dagger$ and $v(r)\simeq v_f$. Thus,
  in our model MOND is just an effective phenomenon of fuzzy DM.
 Therefore,  fuzzy DM can explain the apparent successes of both of CDM and MOND, because
  it acts as CDM at super-galactic scales and as an {\it effective} MOND at galactic scales
  due to the finite length scale $\xi$.
  The mass discrepancy-acceleration relation (MDAR) also appears ~\cite{McGaugh:2004aw},
  because $M_{tot}(r)/M_b(r)=g_{obs}(r)/g_b(r)\simeq \sqrt{g^\dagger/g_b}$, where
  $M_{tot}(r)$ is the total mass enclosed within $r$.
We now understand how RAR  behaves in our model in two extreme limits where
$g_b\gg g^\dagger$ or $g_b\ll g^\dagger$.
An approximate  function
 linking the two limits for RAR  is $g_{obs}=g_b+\sqrt{g^\dagger g_b}$, which is  a simple sum of
 $g_b$ and $g_d\simeq\sqrt{g^\dagger g_b}$ in Eq. (\ref{gobs}) (See Fig. 2).
BTFR and RAR in our model can  have small scatter because these relations
are from the dynamical equilibrium condition
rather than from forming history of galaxies or from baryon physics.

Equation ~(\ref{gobs}) seems to explain some other mysteries  in massive galaxies.
First, for galaxies with flat RCs we can roughly approximate the total density with a cored-isothermal one
$\rho_{obs}\simeq {\sigma^2}/2\pi G(r^2+r^{\dagger 2})\equiv \rho_c/(1+(r/r^\dagger)^2)$ up to a few $r^\dagger$ as an effective core size.
This
leads to an universal surface density of cored galaxies ~\cite{Chan:2013moa}
\beq
\Sigma \simeq \rho_c r^\dagger\simeq \frac{\sigma^2}{2\pi G r^\dagger}
\simeq\frac{g^\dagger}{2\pi G}.
\eeq
Here $\sigma$ is the stellar velocity dispersion and $g^\dagger\simeq \sigma^2/r^\dagger$.
With Eq.~(\ref{gdagger}) this reproduces the observed value ~\cite{2009Natur.461..627G}
$\Sigma=141^{+82}_{-52} M_{\odot} \mbox{pc}^{-2}$
for   $m=1.33^{+.35}_{-.27}\times 10^{-22}e\mbox{V}$ and $\xi=300 \, \mbox{pc}$.
Second, for the isothermal distribution where $g_{obs}\ll g^\dagger$
the wavefunction $\psi$ in the region II should be dynamically
adjusted to satisfy Eq.~(\ref{gobs})
under the small variation of $\rho_b(r)$, which explains the baryon-halo conspiracy for flat RCs~\cite{1985ApJ...293L...7B}.
Finally, we observe that Eq.~(\ref{gobs}) can be rearranged to be an integro-differential equation for $\rho_d(r)$;
\beq
g_b(r)=-g_d\left(\rho_d(r)\right)+\left | \frac{\nabla Q\left(\rho_d(r)\right)}{m} \right |,
\label{gobs2}
\eeq
where $g_b$ plays a role of a source term or
a  boundary condition. A solution $\rho_d(r)$   of
this wave equation
at large $r$ should be such that the right hand side
approaches $ g_b(r)\simeq GM_b/r^2$.
For this solution
details of baryon distribution at central regions except for $M_b$ are not so much relevant.
This explains why $g_d$ and hence $g_{obs}$ are so sensitive to $g_b$
in massive galaxies despite of variety of the galaxies and at the same time insensitive to other visible matter properties like luminosity.

We move to the region III.
In  small dwarf galaxies the spatial size of baryonic matter distribution
is comparable to that of  DM halos, and $M_b$ can not play a role of central
boundary condition as in the region II. Thereby, the arguments related to flat RCs do not hold in this region.
In fuzzy DM model these galaxies
are similar to the ground state (soliton) of boson stars which
has a minimum mass comparable to the quantum Jeans mass.

The mass ($M_{tot}$)-radius ($R$) relation of solitonic core from the boson star theory is
$M_{tot} R=\beta {\hbar^2}/{G m^2}$,
where, for example, the constant $\beta=3.925$ for the half mass radius of DM~\cite{Hui:2016ltb}.
Therefore, using the mass-radius relation the core of DM dominated dwarf galaxies has a typical acceleration
\beq
g_0=\frac{G  M_{tot}}{R^2}\simeq \frac{G^3 m^4 M_{tot}^3}{\beta^2\hbar^4 }
\ge
\frac{G m^4 \gamma^3 M_{J}^3}{\beta^2 \hbar^4},
\eeq
which gives $4.1\times 10^{-12} \mbox{ms}^{-2}$ for
 $m= 1.35\times 10^{-22} e\mbox{V}$ and  $M_{tot}= 10^8 M_\odot$.
 Here we identify  $\gamma M_J$ to be the minimum galaxy mass
from the quantum Jeans mass
\beq
\label{MJ}
M_J(z)
=\frac{\pi^{\frac{13}{4}}}{6}\left(\frac{\hbar}{G^{\frac{1}{2}}   m}\right)^{\frac{3}{2}} \bar{\rho}(z)^\frac{1}{4},
\eeq
where $\gamma\simeq 0.5$ is a numerical constant from numerical studies and
$\bar{\rho}(z)$ is the background matter density at redshift $z$.
Since relevant mass here is the total mass $M_{tot}=M_b+M_d$,
$g_{obs}$ is insensitive to the fraction of baryonic matter
 as long as $M_b \ll M_{tot}$.
This explains the flattening and large scatter of the RAR curve for small dwarf galaxies
where $g_{b}< 10^{-12} \mbox{m/}\mbox{s}^2$ (See Fig.~\ref{rarfig}). Note that $g_0$ has a minimum value from the quantum Jeans mass $M_J$.

Regarding galaxy formation, fuzzy DM has only two free parameters, the particle mass $m$ and  the background matter density $\bar{\rho}(z)$.
If we represent $\xi$ with these parameters, we can fully determine $g^\dagger$ and $A$ from the model. From the boson star mass-radius relation~\cite{sin1,Silverman:2002qx}, a natural candidate for $\xi$ is suggested~\cite{Lee:2015cos,Lee:2008ux} to be
\beq
\label{xi}
\xi=\beta\frac{\hbar^2}{G  M_{tot} m^2}=\frac{3\beta\hbar^{1/2}}{4\pi^{13/4} \gamma (G  m^2 \bar{\rho}(z))^{1/4}},
\eeq
which is about $2\, \mbox{kpc}$ for $M_{tot}=10^8 M_\odot$ and $m=1.3\times 10^{-22}e\mbox{V}$.
This size is somewhat larger than the observed core size $r_c\sim O(10^2)\, \mbox{pc}$ for a massive galaxy, although
the profile of the core is quite similar to the ground state of boson stars.
According to numerical studies with fuzzy DM, the smallness of $r_c$
 is attributed to the nonlocal uncertainty principle applied
to $r_c$ and velocity dispersion $\sigma$, i.e., $r_c \sigma\sim \hbar/m$~\cite{Schive:2014hza}.
More precisely,  $r_c=1.6 a^{1/2}(10^{-22}e\mbox{V/m})(10^9M_\odot/M_h)^{1/3}\mbox{kpc}$, where
$M_h$ is a halo mass~\cite{Schive:2014hza} and $a$ is the scale factor of the universe.
It gives $\xi\simeq r_c=300 \, \mbox{pc}$
for typical halos with $M_h=10^{11}M_\odot$ and $m=1.15\times 10^{-22}\,\mbox{eV}$ at present ($a=1$).
From the $r_c$ formula we expect
  $g^\dagger \propto a^{-3/2} \propto (1+z)^{3/2}$.
Since $r_c$ is a slow function of $M_h$, $\xi$ is almost independent of properties of massive
galaxies such as luminosity. However, in this case, $g^\dagger\sim \xi^{-3}\sim M_h$ depends on the halo mass.
Another possibility is that the self-interaction with $\lambda$ can give a fixed length scale $\xi\sim \sqrt{\lambda} m_p/ m^2$
with the Planck mass $m_p$~\cite{myhalo}.

Our analysis can be easily extended to the Faber-Jackson relation~\cite{1976ApJ...204..668F},
which is an empirical relation $L\propto \sigma^4$
 between the luminosity $L$
   and the central stellar velocity dispersion $\sigma$ of elliptical galaxies.
   If we assume baryon mass to light ratio
    $\Upsilon_b\equiv M_b/L\simeq 3 M_\odot/L_\odot$ is almost constant for elliptical galaxies ~\cite{BT1}
   and $\sigma\sim v_f$ , BTFR in Eq. (\ref{BTF}) implies
   \beq
   L=\frac{M_b}{\Upsilon_b} \simeq \frac{ 34.16 \sigma^4}{\Upsilon_b}
   \left(\frac{m}{10^{-22}\, e\mbox{V}}\right)^{2}
\left(\frac{\xi}{300 \, \mbox{pc} }\right)^3  M_\odot /(\mbox{km/s})^4,
   \eeq
   which is comparable to the observed value
   $ L\simeq 10 L_\odot \sigma^4/(\mbox{km/s})^4$ \cite{1976ApJ...204..668F}.
   Due to differences in $\Upsilon_b$  for individual galaxies,
   we expect larger scatter in the Faber-Jackson relation than in BTFR as observed.

In our simple model with fuzzy DM $g^\dagger$ are not so universal. Interestingly,
a recent observation implies dwarf disc spirals and  Low Surface
Brightness galaxies have different RAR curves and $g^\dagger$ ~\cite{DiPaolo:2018mae}.
There are many studies on the characteristic mass and length scale  in SFDM models, however
little attention has been given to the characteristic acceleration so far ~\cite{Urena-Lopez:2017tob}.
The acceleration scale of fuzzy DM related to the scaling laws such as BTFR and  Faber-Jackson relations
can play an important role in evolution of galaxies and deserves further  studies.
These  relations and observed MOND-like phenomenon in galaxies
seem to add another support for fuzzy DM.
In theoretical point of view, the value of $\xi$ for galaxy mass scale  $M_{c}$  is almost the same as the crossover distance due to dark matter in quantum theory of gravity \cite{LeeYang}.
This work will provide an avenue in understanding  the nature of quantum gravity because the properties of characteristic length scale is related to those in emergent quantum gravity.

\acknowledgments
\vskip 5.4mm
Authors are thankful to Scott Tremaine for helpful comments.


\begin{thebibliography}{10}

\bibitem{McGaugh:2000sr}
S.~S. McGaugh, J.~M. Schombert, G.~D. Bothun, and W.~J.~G. de~Blok, Astrophys.
  J. {\bf 533},  L99  (2000).

\bibitem{2041-8205-816-1-L14}
F. Lelli, S.~S. McGaugh, and J.~M. Schombert, The Astrophysical Journal Letters
  {\bf 816},  L14  (2016).

\bibitem{PhysRevLett.117.201101}
S.~S. McGaugh, F. Lelli, and J.~M. Schombert, Phys. Rev. Lett. {\bf 117},
  201101  (2016).

\bibitem{Ludlow:2016qzh}
A.~D. Ludlow {\it et~al.}, Phys. Rev. Lett. {\bf 118},  161103  (2017).

\bibitem{Trippe:2014hja}
S. Trippe, Z. Naturforsch. {\bf A69},  173  (2014).

\bibitem{1983ApJ...270..365M}
M. {Milgrom}, \apj {\bf 270},  365  (1983).


\bibitem{Milgrom:1992hr}
  M.~Milgrom,
  Annals Phys.\  {\bf 229} (1994) 384


\bibitem{Dodelson:2011qv}
S. Dodelson, Int. J. Mod. Phys. {\bf D20},  2749  (2011).

\bibitem{Salucci:2002nc}
P. Salucci, F. Walter, and A. Borriello, Astron. Astrophys. {\bf 409},  53
  (2003).

\bibitem{navarro-1996-462}
J.~F. Navarro, C.~S. Frenk, and S.~D.~M. White, \apj {\bf 462},  563  (1996).

\bibitem{deblok-2002}
W.~J.~G. {de Blok}, A. Bosma, and S.~S. McGaugh, astro-ph/0212102  (2002).

\bibitem{crisis}
A. Tasitsiomi, International Journal of Modern Physics D {\bf 12},  1157
  (2003).

\bibitem{1983PhLB..122..221B}
M.~R. {Baldeschi}, G.~B. {Gelmini}, and R. {Ruffini}, Physics Letters B {\bf
  122},  221  (1983).

\bibitem{1989PhRvA..39.4207M}
M. {Membrado}, A.~F. {Pacheco}, and J. {Sa{\~n}udo}, \pra {\bf 39},  4207
  (1989).

\bibitem{sin1}
S.-J. Sin, Phys. Rev. {\bf D50},  3650  (1994).

\bibitem{myhalo}
J.-W. Lee and I.-G. Koh, Phys. Rev. {\bf D53},  2236  (1996).

\bibitem{0264-9381-17-1-102}
F.~S. Guzman and T. Matos, Classical and Quantum Gravity {\bf 17},  L9  (2000).

\bibitem{Fuzzy}
W. Hu, R. Barkana, and A. Gruzinov, Phys. Rev. Lett. {\bf 85},  1158  (2000).

\bibitem{Lee:2017qve}
J.-W. Lee, EPJ Web Conf. {\bf 168},  06005  (2018).

\bibitem{2011PhRvD..84d3531C}
P.-H. {Chavanis}, \prd {\bf 84},  043531  (2011).

\bibitem{Hui:2016ltb}
L. Hui, J.~P. Ostriker, S. Tremaine, and E. Witten, Phys. Rev. {\bf D95},
  043541  (2017).

\bibitem{2014ASSP...38..107S}
A. {Su{\'a}rez}, V.~H. {Robles}, and T. {Matos}, Astrophysics and Space Science
  Proceedings {\bf 38},  107  (2014).

\bibitem{2014MPLA...2930002R}
T. {Rindler-Daller} and P.~R. {Shapiro}, Modern Physics Letters A {\bf 29},
  30002  (2014).

\bibitem{2014PhRvD..89h4040H}
T. {Harko}, \prd {\bf 89},  084040  (2014).

\bibitem{2014IJMPA..2950074H}
K. {Huang}, C. {Xiong}, and X. {Zhao}, International Journal of Modern Physics
  A {\bf 29},  50074  (2014).

\bibitem{Marsh:2015xka}
D.~J.~E. Marsh, Phys. Rept. {\bf 643},  1  (2016).

\bibitem{PhysRevD.35.3640}
R. Friedberg, T.~D. Lee, and Y. Pang, Phys. Rev. D {\bf 35},  3640  (1987).

\bibitem{Strigari:2008ib}
L.~E. Strigari {\it et~al.}, Nature {\bf 454},  1096  (2008).

\bibitem{2014NatPh..10..496S}
H.-Y. {Schive}, T. {Chiueh}, and T. {Broadhurst}, Nature Physics {\bf 10},  496
   (2014).

\bibitem{Bhattacharjee:2013exa}
P. Bhattacharjee, S. Chaudhury, and S. Kundu, Astrophys. J. {\bf 785},  63
  (2014).

\bibitem{2017ApJ...840...92L}
P. {Lang} {\it et~al.}, \apj {\bf 840},  92  (2017).

\bibitem{Lelli:2016zqa}
F. Lelli, S.~S. McGaugh, and J.~M. Schombert, Astron. J. {\bf 152},  157
  (2016).

\bibitem{2017ApJ...836..152L}
F. {Lelli}, S.~S. {McGaugh}, J.~M. {Schombert}, and M.~S. {Pawlowski}, \apj
  {\bf 836},  152  (2017).

\bibitem{Bar:2018acw}
N. Bar, D. Blas, K. Blum, and S. Sibiryakov, Phys. Rev. {\bf D98},  083027
  (2018).

\bibitem{Schunck:1998nq}
F.~E. Schunck, astro-ph/9802258  (1998).

\bibitem{Guzman:1999ft}
F.~S. Guzman, T. Matos, and H.~B. Villegas, Astron. Nachr. {\bf 320},  97
  (1999).

\bibitem{Marsh:2015wka}
D.~J.~E. Marsh and A.-R. Pop, Mon. Not. Roy. Astron. Soc. {\bf 451},  2479
  (2015).

\bibitem{2018MNRAS.478.2686C}
J.~H.~H. {Chan}, H.-Y. {Schive}, T.-P. {Woo}, and T. {Chiueh}, \mnras {\bf
  478},  2686  (2018).

\bibitem{1538-3881-143-2-40}
S.~S. McGaugh, The Astronomical Journal {\bf 143},  40  (2012).

\bibitem{Bray:2014dca}
H.~L. Bray and A.~S. Goetz, arXiv:1409.7347  (2014).

\bibitem{2018arXiv180301849W}
C. {Wheeler}, P.~F. {Hopkins}, and O. {Dor{\'e}},   arXiv:1803.01849  (2018).

\bibitem{McGaugh:2004aw}
S.~S. McGaugh, Astrophys. J. {\bf 609},  652  (2004).

\bibitem{Chan:2013moa}
M.~H. Chan, Phys. Rev. {\bf D88},  103501  (2013).

\bibitem{2009Natur.461..627G}
G. {Gentile}, B. {Famaey}, H. {Zhao}, and P. {Salucci}, \nat {\bf 461},  627
  (2009).

\bibitem{1985ApJ...293L...7B}
J.~N. {Bahcall} and S. {Casertano}, \apjl {\bf 293},  L7  (1985).

\bibitem{Silverman:2002qx}
M. Silverman and R.~L. Mallett, Gen.Rel.Grav. {\bf 34},  633  (2002).

\bibitem{Lee:2015cos}
J.-W. Lee, Phys. Lett. {\bf B756},  166  (2016).

\bibitem{Lee:2008ux}
J.-W. Lee, Phys. Lett. {\bf B681},  118  (2009).

\bibitem{Schive:2014hza}
H.-Y. Schive {\it et~al.}, Phys. Rev. Lett. {\bf 113},  261302  (2014).

\bibitem{1976ApJ...204..668F}
S.~M. {Faber} and R.~E. {Jackson}, \apj {\bf 204},  668  (1976).

\bibitem{BT1}
J. Binney and S. Tremaine, {\em Galactic Dynamics}, {\em Princeton Series in
  Astrophysics}, 1st  ed. (Princeton University Press, ADDRESS, 1987).

\bibitem{DiPaolo:2018mae}
  C.~Di Paolo, P.~Salucci and J.~P.~Fontaine,
  Astrophys.\ J.\  {\bf 873}, no. 2, 106 (2019).

\bibitem{Urena-Lopez:2017tob}
L.~A. Urena-Lopez, V.~H. Robles, and T. Matos, Phys. Rev. {\bf D96},  043005
  (2017).

\bibitem{LeeYang}
J. Lee and H. S. Yang, Dark Energy and Dark Matter in Emergent Gravity, [arXiv:1709.04914].


\end{thebibliography}

\end{document}